\documentclass[preprint2]{aastex}

\usepackage{natbib}
\usepackage{graphicx}

\shorttitle{Confirming a Seismologically Inferred Coronal Temperature}
\shortauthors{M.S. Marsh and R.W. Walsh}

\begin{document}

\title{Using HINODE/EIS to Confirm a Seismologically Inferred Coronal Temperature}

\author{M. S. Marsh and R. W. Walsh}
\affil{Jeremiah Horrocks Institute for Astrophysics \& Supercomputing, University of Central Lancashire, Preston, PR1
2HE, UK}
\email{mike.s.marsh@gmail.com}

\begin{abstract}
The Extreme-Ultraviolet Imaging Spectrometer on board the HINODE satellite is used to examine the loop system described in \cite{me09} by applying spectroscopic diagnostic methods. A simple isothermal mapping algorithm is applied to determine where the assumption of isothermal plasma may be valid, and the emission measure locii technique is used to determine the temperature profile along the base of the loop system. It is found that, along the base, the loop has a uniform temperature profile with a mean temperature of $0.89 \pm 0.09$~MK which is in agreement with the temperature determined seismologically in \cite{me09}, using observations interpreted as the slow magnetoacoustic mode. The results further strengthen the slow mode interpretation, propagation at a uniform sound speed, and the analysis method applied in \cite{me09}. It is found that it is not possible to discriminate between the slow mode phase speed and the sound speed within the precision of the present observations. 
\end{abstract}


\section{Introduction}\label{sect_intro}
The temperature of the plasma within coronal loops has traditionally been determined using spectroscopic methods such as emission line ratios \citep{phi08} and emission measure locii techniques \citep{jor87, lan02, del02}. In \cite{me09}, the temperature along a coronal loop structure was seismologically determined using Solar Terrestrial Relations Observatory/Extreme-Ultraviolet Imager \citep[STEREO/EUVI,][]{wue04} observations of wave propagation along a loop system. These waves were interpreted as manifestations of the slow magnetoacoustic mode. The stereoscopic observations were used to derive the propagation geometry with an inclination of ${37 \pm 6} ^{\circ}$ to the local normal and true coronal slow mode phase speed of $132 \pm 9$ km~s$^{-1}$. Thus, the sound speed was then used to infer the plasma temperature of $0.84 \pm 0.12$~MK. 

\cite{me09} was the first direct measurement of the slow mode speed within a coronal loop and inference of the loop plasma temperature using this technique. The work presented here aims to provide an independent observational test of those results and conclusions. The plasma temperature measured using spectroscopic emission line diagnostics is compared to the result obtained using the seismological technique, and confirmed to be in agreement.

\section{Observations}
The observations were conducted on 2008 January 10, as part of the Joint Observing Program (JOP) 200 - `Multi-point, High Cadence EUV Observations of the Dynamic Solar Corona'. Using the Extreme-Ultraviolet Imaging Spectrometer \citep[EIS,][]{cul07} on the HINODE satellite. The HINODE/EIS observations complement the stereoscopic STEREO/EUVI observations with spectroscopic observations from a third point, viewed along the Sun-Earth line. The EIS observations consist of a rastered image using the 2\arcsec slit at $90$ positions with 25~s exposures to build up a $180\arcsec\times512\arcsec$ rastered image from 18:07:32~UT until 18:47:52~UT. The raster study includes data for 24 emission line windows with a width of 48 pixels and a wavelength dispersion of 0.0223~\AA/pixel. 

\begin{figure}[t]
\epsscale{0.8}
\centering
\plotone{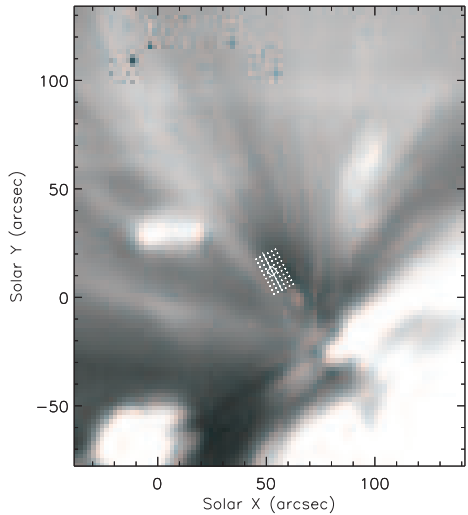}
\caption{\ion{Fe}{12} 195\mbox{\AA} intensity image of the coronal loop system analyzed in \cite{me09}. The loop intensity is extracted from the loop system along the indicated path. The EM locii method is applied using the mean intensity along the path to determine the temperature profile as a function of distance.}
\label{fig1}
\end{figure}

The active region loop system discussed in \cite{me09} is analyzed here, where the loop footpoints are centered on solar $x,y$ coordinates (60\arcsec, 0\arcsec) in the Heliocentric-Cartesian reference frame of the HINODE spacecraft. Figure~\ref{fig1} shows the loop system viewed in emission from \ion{Fe}{12} 195\mbox{\AA}.

\section{Analysis}
\subsection{Preparation of the data}
The EIS data are calibrated using EIS\_PREP within the {\it Solarsoft} database, with standard corrections for dark current, cosmic rays, hot/warm pixels, dusty pixels and an absolute calibration is applied to obtain the data in units of (ergs cm$^{-2}$ s$^{-1}$ s$r^{-1}$ \AA$^{-1}$).

Every two pixels are binned along the $y$ axis to increase the signal to noise ratio within the data, resulting in $2\arcsec \times 2\arcsec$ pixels.
The emission line profiles are then fit with multiple Gaussians using the {\it Solarsoft} routine EIS\_AUTO\_FIT\_GEN. The effects on the line centroids due to the tilt of the EIS slit and the orbital variation are corrected. 

\subsection{EM locii}
\begin{figure}[t]
\epsscale{1.}
\centering
\plotone{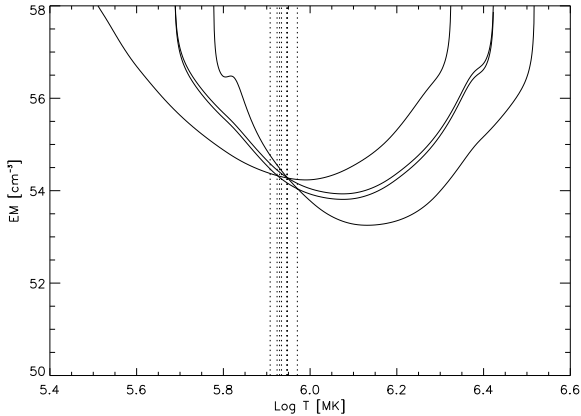}
\caption{Example of the EM locii curves for pixel [46,44]. The dashed lines indicate the points of intersection between each of the curves.}
\label{fig2}
\end{figure}

The temperature of the plasma is investigated using the emission measure locii technique \citep[see][and references within]{jor87, del02, lan02}. If the EM locii curves intersect at a single point (see Figure~\ref{fig2}), it may be assumed that the plasma is isothermal, and the point of intersection used to estimate the plasma temperature and emission measure. To minimize the uncertainties introduced by the elemental abundances, emission lines from the same ion are analyzed. In the case of this data set, the available emission lines of Iron are: \ion{Fe}{10} 184.0\mbox{\AA}, \ion{Fe}{11} 188.23\mbox{\AA}, \ion{Fe}{11} 188.30\mbox{\AA}, \ion{Fe}{12} 186.89\mbox{\AA} and \ion{Fe}{12} 195.12\mbox{\AA}. 


The emission measure locii technique is applied by over plotting the emission measure curves for each line given by the emission measure:

\begin{equation}\label{eqn1}
EM(T)=\frac{4\pi d^{2}I}{G(T)},
\end{equation}

where $I$ is the intensity of the emission line, $d$ is the distance between the emission source and the observer, and $G(T)$ is the contribution function for a particular ionization state. The contribution functions are calculated here using the \mbox{\ion{Fe}{0}} ionization equilibrium of \cite{arn92} and coronal abundances of \cite{fel92a}. To reduce the contamination effect of emission from background plasma, the background emission is estimated and subtracted from the data by subtracting the intensity of pixel [14,22], which is observed to be within a low emission region in all the lines and is adjacent to the loop system. 

\subsection{Isothermal map}
\begin{figure}[t]
\epsscale{.8}
\centering
\plotone{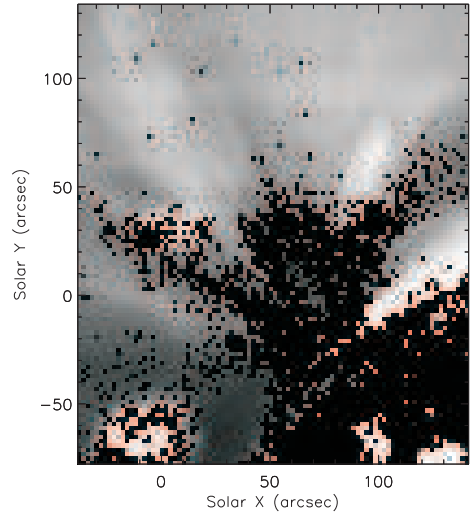}
\caption{Map of isothermal pixels defined by applying the intersection temperature variance algorithm, where isothermal pixels, marked as black, are overplotted on the \ion{Fe}{12} 195\mbox{\AA} intensity image from Figure~\ref{fig1}.}
\label{fig3}
\end{figure}

The emission measure locii technique is based on the assumption of an isothermal plasma. To determine regions where this assumption may be appropriate, a simple algorithm is applied to classify candidate isothermal pixels. Figure~\ref{fig2} shows an example of the EM locii curves for a single pixel within the loop system. The points of intersection between the curves are determined, indicated by dashed lines in Figure~\ref{fig2}. The variance of the temperature values for the intersections is used to classify a pixel as isothermal or not isothermal. Pixels with an intersection temperature variance of Log $T < 0.004$ are defined to be isothermal. Figure~\ref{fig3} shows a map of isothermal pixels defined by applying the algorithm to all pixels within the data. The isothermal map indicates that the region at the base of the loop system is defined as isothermal by the algorithm, thus suggesting that it is reasonable to apply the emission measure locii technique to estimate the temperature at the base of these loops. 

\begin{figure}[t]
\epsscale{1.}
\centering
\plotone{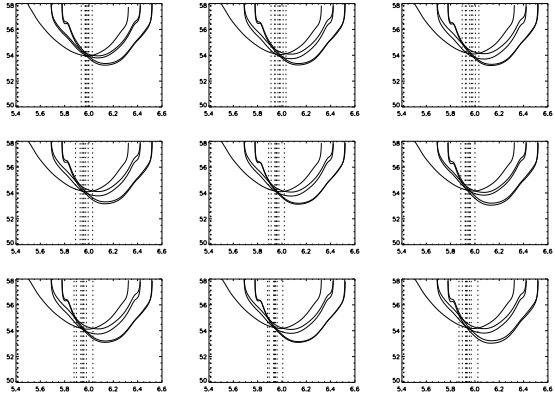}
\caption{EM locii curves calculated using the mean intensity of each cross-section along the path shown in Figure~\ref{fig1}, where Log $T$ [MK] is along the abscissa and emission measure [cm$^{-3}$] is along the ordinate. }
\label{fig4}
\end{figure}

\begin{figure}[t]
\epsscale{1.}
\centering
\plotone{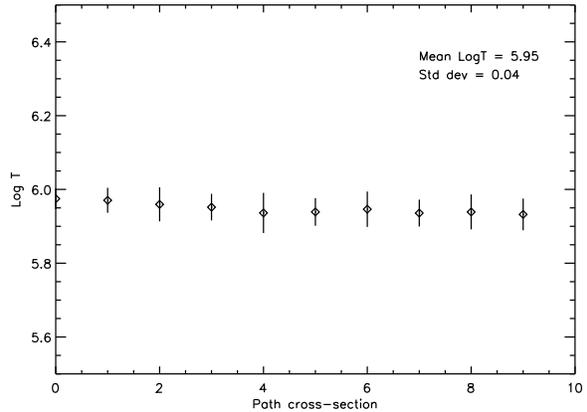}
\caption{Temperature profile as a function of distance along the path indicated in Figure~\ref{fig1}, derived using the EM locii method and intersection temperature standard deviation.}
\label{fig5}
\end{figure}

\section{Results}
\subsection{Temperature profile}
The temperature profile of the loop system is investigated by applying the EM locii method along the loops. The intensity profile of the loop system is extracted by defining a path parallel to the loops with a width of 6 pixels (Figure~\ref{fig1}). In each cross-section, perpendicular to the path axis, the mean intensity of the pixels is calculated. Thus the mean intensity is determined as a function of distance along the loops. The emission measure locii curves are calculated using the mean intensity of each cross-section along the path as shown in Figure~\ref{fig4}. The temperature profile along the loop path is derived using the EM locii curves as shown in Figure~\ref{fig5}, where the temperature error is calculated using the standard deviation of the intersection temperature values. The temperature profile suggests a uniform temperature as a function of distance along the loop within the errors. The mean temperature along the whole length of the temperature profile is Log $T =5.95 \pm 0.04$~K or equivalently $T =0.89 \pm 0.09$~MK, compared to the seismological result of $0.84 \pm 0.12$~MK obtained in \cite{me09}.

\section{Conclusions}
The results presented here determine, spectroscopically, the temperature of the active region loop system presented in \cite{me09}, along which slow magnetoacoustic waves were found to propagate. The seismological technique, applied in \cite{me09}, derived a plasma temperature of  $0.84 \pm 0.12$~MK. This temperature is independently confirmed, here, using the emission measure locii technique, with a derived temperature of $T =0.89 \pm 0.09$~MK, and is consistent with the \cite{me09} results. 

The agreement between the results validates the technique applied in \cite{me09}, and further strengthens the slow magnetoacoustic mode interpretation of the observed waves. The consistency between the two independent estimates of the temperature also suggests that the assumption of an isothermal plasma in the loop system is valid. The temperature measured at the base of the loop system shown in Figure~\ref{fig5} displays a uniform temperature profile. This result is also in agreement with the EUVI observations presented in \cite{me09}, where the observed waves have a constant phase speed as a function of distance. Thus, indicating a uniform temperature profile along the base of the loop system, at least to the extent of where the waves are observed. 

\cite{wan09} recently investigated waves similar to those reported in \cite{me09} and found consistent results. Using HINODE/EIS observations interpreted as slow magnetoacoustic waves at the footpoint of a coronal loop, they use the Doppler shift amplitude to estimate the loop inclination and estimate a temperature of $0.7 \pm 0.3$~MK. Again, this is consistent with the results of the spectroscopic diagnostics applied here. 

In \cite{me09}, it was suggested that it may be possible to use the propagating slow mode to measure the coronal magnetic field strength. The HINODE/EIS observations presented here indicate that, at least within the current observational diagnostic precision, at the temperature and density of this region, it is not possible to discriminate the slow mode tube speed from the sound speed to make such a measurement. This may be possible in higher temperature structures where, due to its dependence on $T$, the sound speed would be expected to have a greater divergence from the tube speed. This may be a possible achievable goal of the Solar Orbiter mission, with high resolution observations of wave propagation in different structures.

\acknowledgments
This research is supported by the Science and Technology Facilities Council (STFC) under grant number ST/F002769/1. Hinode is a Japanese mission developed and launched by ISAS/JAXA, with NAOJ as domestic partner and NASA and STFC (UK) as international partners. It is operated by these agencies in co-operation with ESA and NSC (Norway).
M.S. Marsh would like to acknowledge the encouragement of L.E. Marsh.

{\it Facilities:} \facility{HINODE (EIS)}.

\bibliographystyle{apj}

\end{document}